\documentclass[11pt]{amsart}
\usepackage{amsxtra,amssymb,amsmath,amsthm,amsbsy}
\usepackage{graphicx}

\usepackage[T1]{fontenc}
\usepackage[french]{babel}

\usepackage{graphicx}

\begin{document}
\title{Note sur la Détermination des paramètres du Modèle de Bursa-Wolf}

\author{ Abdelmajid Ben Hadj Salem}\thanks{Ingénieur Général, Chef de la Division de la Coopération Technique et le Développement des Ressources Humaines. \\
Office de la Topographie et du Cadastre (OTC), \\
 BP 156, 1080 Tunis Cedex, Tunisie\\
 Email: benhadjsalema@yahoo.co.uk}
 \begin{center}
\Large
\textsc\textbf{MINISTERE DE L'EQUIPEMENT}\\
\textsc{Office de la Topographie et du Cadastre }\\

\vspace{4cm}

\LARGE
\textsc{\textbf{Note sur la Détermination des paramètres du Modèle de Bursa-Wolf}}\\
[0.5\baselineskip]
{Par \\ Abdelmajid BEN HADJ SALEM}\\
\vspace{0.5cm}
\normalsize
\textsc{Ingénieur Général à l'Office de la Topographie et du Cadastre}\\

\vspace{1cm}
\textsc{Novembre 2011}
\\ 

\vspace{1cm}
\textsc{Version 1.}\\

\vspace{4cm}
\textsc{Office de la Topographie et du Cadastre \\
www.otc.nat.tn}
\end{center}

\maketitle 


\section{ Introduction}

        Avec le développement de la technologie de positionnement spatial (GPS, GLONASS, Galileo, ComPass), laquelle fournit à l'utilisateur sa position $(X,Y,Z)$ tridimensionnelle dans un système géocentrique mondial donné, par exemple pour la technologie GPS c'est le système  dit $WGS84$ (World Geodetic System 1984), il est nécessaire de savoir la transformation de passage du système géodésique mondial au système géodésique national ou local. Nous présentons ci-après en détail comment déterminer manuellement les paramètres du modèle de Bursa-Wolf de transformations de passage entre les systèmes géodésiques.
\\
        
 Nous utilisons par la suite les notations suivantes :
 
-	$(X_1,Y_1,Z_1)$  les coordonnées cartésiennes 3D dans le système géocentrique $(O,X_1,Y_1,Z_1)$(système 1),

-	$(X_2,Y_2,Z_2)$ les coordonnées cartésiennes 3D dans le système local (système 2) $(O',X_1,Y_1,Z_1)$.

\section{Le Modèle de  BURSA - WOLF}
Ce modèle s'écrit sous la forme vectorielle :
\begin{equation}
	\textbf{\textit{X}}_2 = \textbf{\textit{T}} + ( 1+m).R(rx,ry,rz).\textbf{\textit{X}}_1  \label{yy1}
\end{equation}
où:

 - $\textbf{\textit{X}}_2$  est le vecteur de composantes $(X_2,Y_2,Z_2)^T$ , $T$ désigne transposée,
  
 - $\textbf{\textit{T}}=\textbf{\textit{O'O}}$ est le vecteur translation de composantes $(T_X,T_Y,T_Z)^T$  entre les systèmes 1 et 2,
       
 - $1+m$ est le facteur d'échelle entre les 2 systèmes,
      
 - $R(rx,ry,rz)$ est la matrice de rotation $(3\times3)$ pour passer du système 1 au système 2,
      
 -  $\textbf{\textit{X}}_1$ est le vecteur de composantes $(X_1,Y_1,Z_1)^T$.
 \\
        
En développant \eqref{yy1}, on obtient:
      \begin{equation}
	\begin{pmatrix}
	X_2 \\	  Y_2 \\ 	  Z_2
	\end{pmatrix}=	
	\begin{pmatrix} 
	T_X \\ T_Y \\ T_Z
	\end{pmatrix}+(1+m)
	\begin{pmatrix}
	1     &  -rx    &   ry  \\
 	rx  &   1     & -rz \\
	-ry    &   rz &  1   \\
\end{pmatrix} 
	\begin{pmatrix} 
	X_1 \\ Y_1 \\ Z_1
	\end{pmatrix}\label{yy2}
\end{equation}
avec $(rx,ry,rz)$ les rotations comptées positivement dans le sens contraire des aiguilles d'une montre. Comment déterminer les paramètres modèle \eqref{yy1}?

\section{Détermination de l'échelle $1+m$}
On suppose donné un ensemble de points $Pi$ pour $i=1,n$ connus dans les deux systèmes $S_1$ et $S_2$. On écrit l'équation \eqref{yy1} pour deux points $Pj$ et $Pk$, d'où:
\begin{align}
& \textbf{\textit{X}}(Pj)_2 = \textbf{\textit{T}} + ( 1+m).R(rx,ry,rz).\textbf{\textit{X}}(Pj)_1  \label{yy3} \\
&	\textbf{\textit{X}}(Pk)_2 = \textbf{\textit{T}} + ( 1+m).R(rx,ry,rz).\textbf{\textit{X}}(Pk)_1  \label{yy4} 
\end{align}
Par différence, on obtient :
\begin{equation}
	(\textbf{Pj}\textbf{Pk})_2=(1+m).R(rx,ry,rz).(\textbf{Pj}\textbf{Pk})_1  \label{yy5} 
\end{equation}
On prend la norme des deux membres de \eqref{yy5} et que $1+m > 0$:
\begin{equation}
\|(\textbf{Pj}\textbf{Pk})_2\|=\|(1+m).R(rx,ry,rz).(\textbf{Pj}\textbf{Pk})_1\|=(1+m)\|R(rx,ry,rz).(\textbf{Pj}\textbf{Pk})_1 \|   \label{yy6} 
\end{equation}
Comme $R$ est une matrice de rotation, donc son application à un vecteur est une isométrie, c'est-à-dire qu'elle laisse invariant la norme ou la longueur du vecteur soit:
\begin{equation}
	\|R.\textbf{\textit{X}}\|=\|\textbf{\textit{X}}\| \quad \forall \,\textbf{\textit{X}} \label{yy7}
\end{equation}
On a donc:
\begin{equation}
\|(\textbf{Pj}\textbf{Pk})_2\|=(1+m)\|(\textbf{Pj}\textbf{Pk})_1 \|   \label{yy8} 
\end{equation}
Soit:
\begin{equation}
	1+m=\frac{1}{N}\sum^N\frac{\|(\textbf{Pj}\textbf{Pk})_2\|}{\|(\textbf{Pj}\textbf{Pk})_1\|} \label{yy9}
\end{equation}
$N$ désigne le nombre de couples de points $PjPk, j\neq k$.
\section{Détermination des rotations $(rx,ry,rz)$}
Connaissant $(1+m)$, pour un couple de points $Pj,Pk$, on a :
\begin{equation}
	(\textbf{Pj}\textbf{Pk})_2=(1+m).R(rx,ry,rz).(\textbf{Pj}\textbf{Pk})_1  \nonumber  
\end{equation}
 Détaillons la matrice $R$:
 \begin{equation}
	R=	\begin{pmatrix}
	1     &  -rx    &   ry  \\
 	rx  &   1     & -rz \\
	-ry    &   rz &  1   \\
\end{pmatrix} =
	\begin{pmatrix}
	1     & 0    &   0  \\
 	0  &   1     & 0 \\
	0    &   0 &  1   \\
\end{pmatrix} +
	\begin{pmatrix}
	0     &  -rx    &   ry  \\
 	rx  &   0     & -rz \\
	-ry    &   rz &  0   \\
\end{pmatrix}=I_3+Q \label{yy10}
\end{equation}
avec $I_3$ la matrice Unité et $Q$ la matrice:
\begin{equation}
	Q=
	\begin{pmatrix}
	0     &  -rx    &   ry  \\
 	rx  &   0     & -rz \\
	-ry    &   rz &  0   \\
\end{pmatrix}   \label{yy11}
\end{equation}
Alors l'équation \eqref{yy5} devient:
\begin{equation}
	(\textbf{Pj}\textbf{Pk})_2=(1+m).(I_3+Q(rx,ry,rz)).(\textbf{Pj}\textbf{Pk})_1  \label{yy12} 
\end{equation}
Soit comme $m << 1$ et $m² << 1$:
\begin{equation}
Q(rx,ry,rz).(\textbf{Pj}\textbf{Pk})_1=(1-m).(\textbf{Pj}\textbf{Pk})_2-(\textbf{Pj}\textbf{Pk})_1 \label{yy13} 
\end{equation}
En posant:
\begin{equation}
	(\textbf{Pj}\textbf{Pk})_2=
\begin{pmatrix}
\Delta X'_{jk} \\
\Delta Y'_{jk} \\
\Delta Z'_{jk}
\end{pmatrix};\quad 	(\textbf{Pj}\textbf{Pk})_1=
\begin{pmatrix}
\Delta X_{jk} \\
\Delta Y_{jk} \\
\Delta Z_{jk}
\end{pmatrix};\quad v=
\begin{pmatrix}
v_1=(1-m)\Delta X'_{jk}-\Delta X_{jk} \\
v_2=(1-m) \Delta Y'_{jk}-\Delta Y_{jk} \\
v_3=(1-m)\Delta Z'_{jk}-\Delta Z_{jk}
\end{pmatrix} \label{yy14}
\end{equation}
Alors, on obtient l'équation:
\begin{equation}
	Q(rx,ry,rz).(\textbf{Pj}\textbf{Pk})_1=v \label{yy15}
\end{equation}
ou encore:
\begin{equation}
	\begin{pmatrix}
	-\Delta Y_{jk} & \Delta Z_{jk} & 0 \\
	\Delta X_{jk} & 0 & -\Delta Z_{jk} \\
	0 & -\Delta X_{jk} & \Delta Y_{jk}
\end{pmatrix}.
\begin{pmatrix}
	rx \\
	ry \\
	rz
\end{pmatrix}=
\begin{pmatrix}
	v_1\\
	v_2\\
	v_3
\end{pmatrix} \label{yy16}
\end{equation}
Or le déterminant de la matrice $Q'$ :
\begin{equation}
Q'=	\begin{pmatrix}
	-\Delta Y_{jk} & \Delta Z_{jk} & 0 \\
	\Delta X_{jk} & 0 & -\Delta Z_{jk} \\
	0 & -\Delta X_{jk} & \Delta Y_{jk}
\end{pmatrix} \label{yy17}
\end{equation}
est nul. Pour passer de cette conséquence, on utilise pour chaque ligne du système \eqref{yy16} un couple de points $ij$ ce qui donne le système:
\begin{equation}
\begin{pmatrix}
	-\Delta X_{jk} & \Delta Z_{jk} & 0 \\
	\Delta X_{lm} & 0 & -\Delta Z_{lm} \\
	0 & -\Delta X_{in} & \Delta Y_{in}
\end{pmatrix}.
\begin{pmatrix}
	rx \\
	ry \\
	rz
\end{pmatrix}=
\begin{pmatrix}
	v_{jk1}\\
	v_{lm2}\\
	v_{in3}
\end{pmatrix} \label{yy18}
\end{equation}
Le système \eqref{yy18} devient résolvable ce qui permet de déterminer les trois rotations $rx,ry$ et $rz$.
\section{Détermination des composantes de la Translation $T$}
Les composantes $Tx,Ty,Tz$ du vecteur translation sont déterminées à partir des coordonnées des points $Pj$ connus dans les deux systèmes à partir de:
\begin{align}
&	Tx_j=X_{2j}-(1+m)(X_{1j}-rxY_{1j}+ryZ_{1j}) \label{yy19} \\
&		Ty_j=Y_{2j}-(1+m)(rxX_{1j}+Y_{1j}-rzZ_{1j}) \label{yy20} \\
&	Tz_j=Z_{2j}-(1+m)(-ryX_{1j}+rzY_{1j}+Z_{1j}) \label{yy21} 
\end{align}
Les composantes $Tx,Ty,Tz$ sont obtenues par une moyenne sur les $N$ points communs à savoir:
\begin{align}
	& Tx=\frac{\sum_j^NTx_j}{N} \label{yy22} \\
	&	Ty=\frac{\sum_j^NTy_j}{N} \label{yy23} \\
&	Tz=\frac{\sum_j^NTz_j}{N} \label{yy24}
\end{align}
\section{Référence}


\textbf{1.}	\textbf{A. Ben Hadj Salem}. (2011). Cours d'initiation au GPS. v1. 32 p.


\end{document}